\documentclass[twocolumn,twoside]{revtex4}
\usepackage{graphicx}
\usepackage{fancyhdr}
\begin{document}

\title{Future High Energy Frontier Colliders}

%

\author{Vladimir Shiltsev}
\affiliation{Fermilab, MS312, PO Box 500, Batavia, IL, 60510, USA }

\begin{abstract}
Colliders have been at the forefront of scientific discoveries in high-energy particle physics since the inception of the colliding beams method in the middle of the 20th century. The field of accelerators is very dynamic and many innovative concepts are currently being considered such future facilities as Higgs factories and energy frontier colliders beyond the LHC. Here we briefly overview leading proposals and studies towards the next generation colliders and discuss their major challenges as well as directions of corresponding accelerator R\&D programs needed to address their cost and performance risks. 
\end{abstract}

\maketitle

\thispagestyle{fancy}


\section{INTRODUCTION}
The needs of modern high energy physics (HEP) call for two types of future accelerator facilities –- Higgs Factories (HF) and Energy Frontier  (EF) colliders. There are four feasible concepts for these machines: linear $e^+e^-$ colliders, circular $e^+e^-$ colliders, $pp/ep$ colliders, and muon colliders. They all have limitations in energy, luminosities, efficiencies, and costs which in turn depend on five basic underlying accelerator technologies: normal-conducting (NC) magnets, superconducting (SC) magnets, NC RF, SC RF and plasma. The technologies are at different level of performance and readiness, cost efficiency and required R\&D. Comprehensive reviews of colliders can be found in \cite{Handbook}, \cite{VS2012}, \cite{BM2016}, \cite{RASTv7}, \cite{SZ}. Below we overview the Higgs factory implementation options, their accelerator physics and technology challenges, readiness, cost and power; possible paths towards the highest energies, how to achieve the ultimate energy and performance, and required R\&D; as well as  promises, challenges and expectations of new acceleration techniques. There is an extended literature for each of the considered future colliders but in this paper, for reasons of brevity, references will be given the corresponding brief Inputs to the 2019 European Particle Physics Strategy Update symposium (EPPSU, May 2019, Granada, Spain) where all the proposals were presented \cite{EPPSU}. More details and references on each of the proposals can be found therein. Table \ref{future} summarizes main parameters of the future facilities. 

\begin{figure}[h]
\centering
\includegraphics[width=80mm]{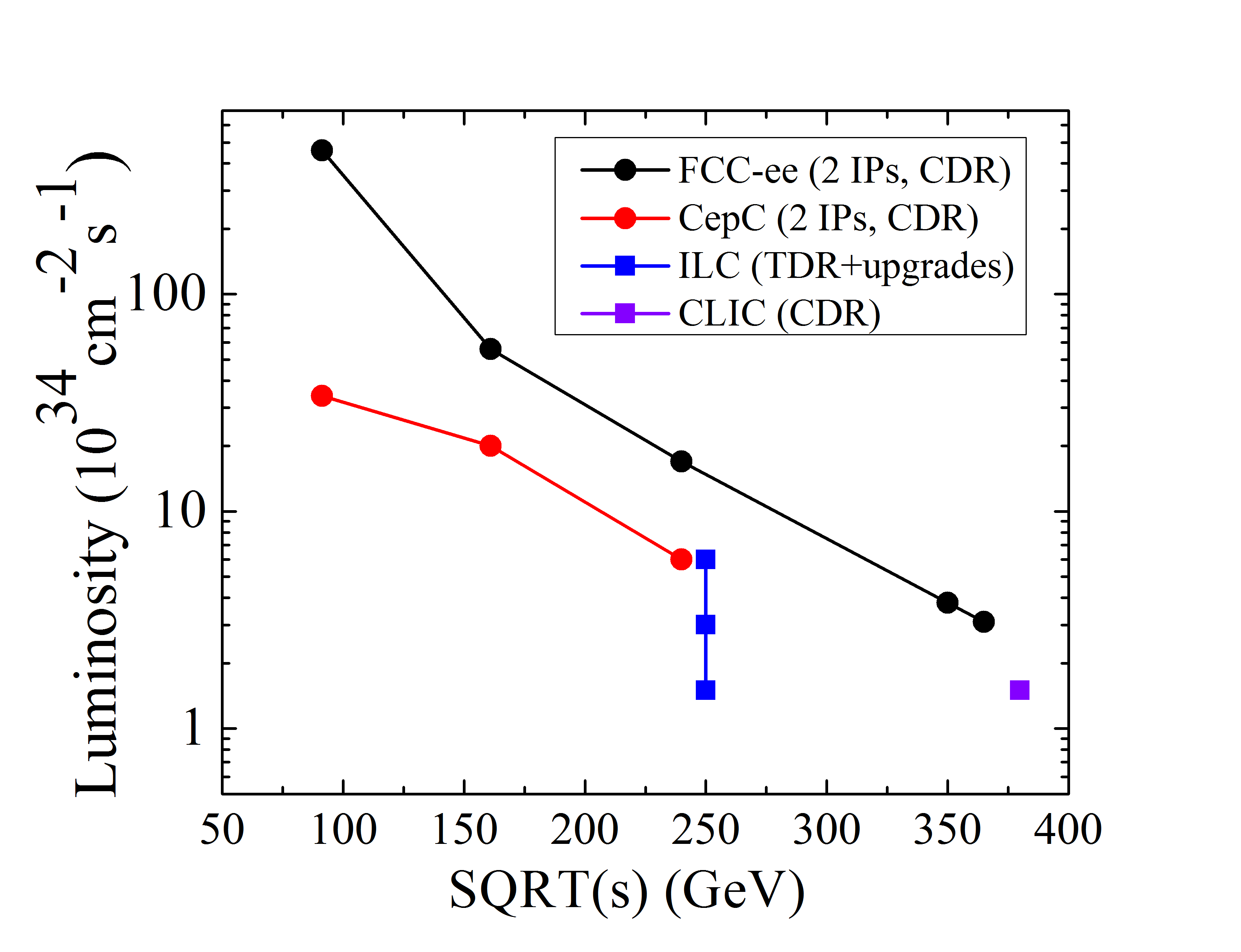}
\caption{Luminosity of the proposed Higgs factories.} 
\label{HFlumis}
\end{figure}

\begin{table*}[t]
\begin{center}
\caption{Main parameters of proposed colliders for high energy particle physics research.}
\begin{tabular}{|l|c|c|c|c|c|c||c|c|}
\hline 
\textbf{Project} & \textbf{Type} & \textbf{Energy} & \textbf{Int. Lumi.} &
\textbf{Oper.Time} & \textbf{Power} & \textbf{Cost} & \textbf{Cost/Lumi.} & \textbf{Lumi./Power} \\
 & &TeV, c.m.e. & ab$^{-1}$ & years & MW & & BCHF/ab$^{-1}$ & ab$^{-1}$/TWh\\
\hline 
\textbf{ILC} & $e^+e^-$ & 0.25 & 2 & 11 & 129 & 4.8-5.3GILCU & 2.65 & 0.24 \\
$L$-upgr. &  &  &  &  & 150-200 & + upgrade &  &  \\
 &  & 0.5 & 4 & 10 & 163(204) & 7.8GILCU & 1.3 & 0.4 \\
 &  & 1 &  &  & 300 & ? &  &  \\
\hline 
\textbf{CLIC} & $e^+e^-$ & 0.38 & 1 & 8 & 168 & 5.9BCHF & 5.9 & 0.12 \\
 &  & 1.5 & 2.5 & 7 & 370 & + 5.1BCHF & 3.1 & 0.16 \\
 &  & 3 & 5 & 8 & 590 & +7.3BCHF & 2.0 & 0.18 \\
\hline
\textbf{CEPC} & $e^+e^-$ & 0.091+0.16 & 16+26 & 4 & 149 & 5 G\$ & 0.27 & 5.1 \\
 &  & 0.24 & 5.6 & 7 & 266 & +? & 0.21 & 0.5 \\
\hline
\textbf{FCC-ee} & $e^+e^-$ & 0.091+0.16 & 150+10 & 4+1 & 259 & 10.5BCHF & 0.065 & 20.5 \\
 &  & 0.24 & 5 & 3 & 282 &  & 0.064 & 0.9 \\
 &  & 0.365(+0.35) & 1.5(+0.2) & 4(+1) & 340 & +1.1 BCHF & 0.07 & 0.15 \\
\hline
\hline
\textbf{LHeC} & $ep$ & 0.06/7 & 1 & 12 & (+100) & 1.75 GCHF & 1.75 & 0.14 \\
\hline
\textbf{HE-LHC} & $pp$ & 27 & 20 & 20 & 220 & 7.2 GCHF & 0.36 & 0.75 \\
\hline
\textbf{FCC-hh} & $pp$ & 100 & 30 & 25 & 580 & 17(+7) GCHF & 0.8
& 0.35 \\
\hline
\textbf{Muon.Coll.} & $\mu\mu$ & 14 & 50$^*$ & 15$^*$ & 230 & 10.7$^*$ GCHF & 0.21$^*$ & 2.4$^*$ \\
\hline
\hline
\end{tabular}
\label{future}
\end{center}
\end{table*}

\section{HIGGS FACTORIES}
There are several well studied approaches to produce copious number of Higgs particles:  $e^+e^-$ linear colliders such as ILC (the EPPSU Input 77) and CLIC (Input 146); $e^+e^-$ circular colliders like FCC-ee (Input 132) and CepC (Input 51) and a $\mu^+\mu^-$ circular HF. Less traditional and yet not that well studied options such as $\gamma\gamma$-colliders are described elsewhere \cite{HF2012} and are not discussed here. The most critical high level requirement for HFs is high luminosity $O(10^{34})$ cm$^{-2}$s$^{-1}$ at the Higgs energy scale. 
Usually, these machines are compared to the LHC which, as an accelerator, is
27 km long, is based on 8 T SC magnets, requires some 150 MW total AC power (out of 200MW for the entire site of CERN which consumes some 1-1.2 TWh of electric energy annually), took about 10 years to build after 10-15 years of the design and development studies, and did cost about 5 BCHF (excluding the cost of the existing LEP tunnel and of the proton injector complex). 

\subsection{Linear Colliders}

The International Linear Collider (ILC, EPPSU Inputs 66 and 77) is about
20 km long, including 5 km of the Final Focus system, it employs 1.3 GHz SRF cavities operating at 2 K and providing 31.5 MV/m of the accelerating gradient. The ILC requires some 130 MW site power when operates at 250 GeV c.m.e. It is estimated to cost 700 BJPY (with $\pm$25\% error, including cost of labor). 

The 380 GeV c.m.e. Compact Linear Collider (CLIC) is 11 km long, its two-beam NC RF accelerator cavities operate at 72 MV/m. Out of the total 168 MW site power some 9MW goes into electron and positron beams. The CLIC cost estimate is 5.9 BCHF with some $\pm$25\% accuracy. 

Luminosity of the linear colliders scales as:
\begin{equation}
L = H_d \Big( \frac{N_e}{\sigma_x} \Big) \Big( N_e N_b f_r \Big) \Big( \frac{1}{\sigma_y} \Big).
\label{LumiHiggs}
\end{equation}
Correspondingly, the luminosity challenges are associated with each of three limiting factors in the parentheses: the first one $(N_e / \sigma_x)$  defines the luminosity spectrum which reaches $\delta E/E$ $\sim$1.5\% in ILC and grows with energy, so some 40\% of  the CLIC luminosity is out of  1\% c.m.e. (due to so called $beamstrahlung$); the second factor is nothing but the total beam current which is limited by the total available RF power, by the beam coherent instability concerns, by challenging positron production via two proposed schemes, etc; and the third one calls for ultra small vertical beam size that in turns requires record small beam emittances from damping rings, stabilization of focusing magnets and accelerating cavities \cite{Shi1}, \cite{Shi2}, 0.1 $\mu$m resolution BPMs in order to obtains the rms beam sizes of 8nm (vertical) and 500nm (horizontal) at the ILC interaction point(IP) and 3nm/150nm in the CLIC IP. 
One should note remarkable progress of the linear collider projects: beam accelerating gradients met the ILC specs of 31.5 MeV/m (at the Fermilab FAST facility in 2017 \cite{BROEMM} and KEK in 2019) and the CLIC specs of 100 MeV/m at the CLEX facility at CERN; final focusing into 40 nm verstical rms beam size has been demonstrated at the ATF2 in KEK in 2016, etc. 

In general, one should consider numerous advantages of the linear $e^+e^-$ colliders as HFs: they are based on mature technologies of NC RF and SRF; their designs are quite mature (ILC has a TDR, CLIC has CDR, there are several beam test facilities); beam polarization of 80\%-30\% in the ILC and 80\% - 0\% in CLIC helps the HEP research; they are expandable to higher energies (ILC to 0.5 and 1 TeV, CLIC to 3 TeV); both have well-organized established international collaborations (LCC) which indicate readiness to start costruction soon; the AC wall plug power of 130-170 MW is less than that of the LHC. 
At the same time one has to pay attention to following factors: the cost of these facilities is 1-1.5 times the LHC cost; the ILC and CLIC luminosity projections are in general lower than that for rings (see Fig.\ref{HFlumis} and discussion below), and luminosity upgrades (such as via two-fold increase of the number bunches $N_b$ and doubling the repetition rate from 5 Hz to 10 Hz in the ILC) will probably come at an additional cost; operation experience with LCs is limited to one past machine (SLAC's SLC), CLIC's two-beam scheme is quite novel (so, klystrons are assumed as a backup RF source option); and the wall plug power may grow beyond that of the LHC for the proposed LC luminosity and energy upgrades. 
 
\subsection{Circular colliders}
The most advanced circular electron-positron Higgs factories designs are CERN's FCC-ee (Input 132) and Chinese CepC (Input 51). Both proposals call for 100 km tunnels to host three rings ($e^-, e^+$ and a fast cycling booster ring), very high SRF power transfer to beams (100 MW in FCC-ee and 60 MW in CepC), that leads to the 
total site power of about 300MW. Cost estimate of the FCC-ee is 10.5 BCHF (plus additional 1.1BCHF for option to operate at the higher $tt$ energy) and 5 to 6 B\$ for the CepC (“less than 6BCHF” cited in the CepC CDR). 

The key accelerator physics challenge of the circular HFs is that the synchrotron radiation power from both beams has to be limited to about $P$=100 MW and the maximum allowed beam current $I$ scales as inverse fourth power of the beam energy:  
\begin{equation}
I = P \Big( \frac{e\rho}{2 C_\gamma E^4} \Big).
\label{SRcurrent}
\end{equation}
Correspondingly, the luminosity scales as the product of the ring radius $\rho$, beam-beam parameter $\xi_y$, beta-function at the IP $\beta_y^*$ and the RF power $P$ and inverse of $E^3$. The beam-beam parameter is limited to about $\xi_y$=0.13 by a new type of beam-beam instability. Beam energy loss per turn due to the synchrotron radiation is about 0.1-5\% (from $Z$ energy to 360 GeV) is significantly more than the energy spread due to beamstrahlung 0.1-0.2\%, but the latter occurs instanteneously at the IPs and the tails of the resulting energy distribution reach $\pm$2.5\% or upto 10 times the rms value. Correspondingly, these tails determine 18 min beam lifetime even in a sophisticated large energy  acceptance $crab-waist$ optics  with $\beta^*_y$ = 0.8-1.6 mm -- that in turn, calls for a full energy fast ramping booster ring. 

The advantages of the circular HFs include quite mature technology of the SRF acceleration, vast experience of other rings suggests lower performance risk; they have higher luminosity and luminosity/cost ratio and can host upto 4 detectors at IPs that could make them sort of {\it EW} (electroweak) factories. 100 km long tunnels can be reused by follow-up future $pp$ colliders. Transverse polarization occurs naturally after about 18 min at the $tt$ energies and can be employed for precise energy calibration $O$(100keV). Very strong and broad {\it Global FCC Collaboration} came up with comprehensive CDR that addresses key design points and indicates possible ca.2039 start (the CepC schedule is more aggressive with the machine start some 7-9 years sooner). Before that, the R\&D program is expected to address several important items, such high efficiency RF sources (e.g. over over 85\% for 400/800 MHz klystrons, up from thecurrent 65\%); high efficiency SRF cavities (to achieve 10-20 MV/m CW gradient and high $Q_0$; or use of new technologies like Nb-on-Cu, Nb$_3$Sn); 
exploration of the crab-waist collision scheme (the Super KEK-B  nanobeams experience will be very helpful in that regard); energy storage and release (so 
energy in magnets can be re-used for more that 20,000 cycles) and on the efficient use of excavated materials (some 10 million cu.m. will need to be taken out of a 100 km tunnel). 

Finally, muon collider HF (Inputs 41, 120, 141) might offer a very economical approach as the luminosity required in $\mu^+\mu^-$ collisions can be about 1/100th of that in  $e^+e^-$ due to large cross-section in $s$-channel; beam energy is also only one half of the $e^+e^-$ case (i.e., 2$\times$63 GeV  for $\mu^+\mu^- \rightarrow H_0$) and, therefore, a small footprint of less than 10 km and low cost of the collider; very small energy spread in non-radiating muon beams  $O$(3 MeV) and low total site power $\sim$200MW. The biggest challenge of all $\mu^+\mu^-$ collider proposals is that the method, though based mostly on conventional technologies, is quite novel conceptually, and some key techniques, such as efficient muon cooling, are still being explored. So, the muon colliders offer great cost savings for future HF and EF machines and substantial R\&D needs to be carried out to prove their practicality to be considered on equal footing with the above mentioned proposals (see also discussion in the next section).  
\begin{figure}[h]
\centering
\includegraphics[width=80mm]{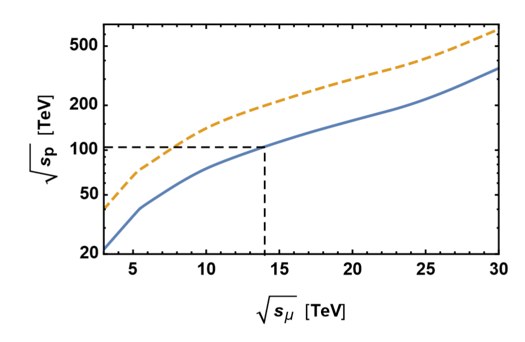}
\caption{Energy reach of muon-muon collisions: the energy at which the proton collider cross-section equals that of a muon collider (from the EPPSU Input 120).} 
\label{MuonProton}
\end{figure}

\section{FUTURE ENERGY FRONTIER FACILITIES}
Key challenges of all EF collider proposals are mostly about having 
affordable cost and, simultaneously, high luminosity. Usually, the 
scale of civil construction grows with beam energy, the cost of accelerator components grows with energy, required site power grows with energy. So, the 
total project cost grows with energy, but, thankfully, not linearly \cite{VScost}. Take the ILC as an example: the costs of 0.25 TeV vs 0.5 TeV vs 1 TeV facilities relate as 0.69 : 1 : 1.67. Still, there are huge financial challenges for collision energies an order of magnitude beyond the LHC's 14 TeV. 

Linear $e^+e^-$ colliders face the most severe challenges: both ILC and CLIC offer staged approach to ultimate energies (1 TeV and 3 TeV c.m.e., respectively), but their lengths grow to 40-50 km, the AC power 
requirements become 300-600 MW, the beamstrahlung leaves only 30-40\% of the luminsity within 1\% of the maximum energy and the project cost grows to 17 B\$ (ILC 1 TeV, TDR) and 18.3 BCHF (CLIC 3 TeV, CDR). 

EF $pp$ colliders such as HE-LHC (Input 133), FCC-hh (Input 136) and SppC (Input 51) require long tunnels (27km, 100km and 100 km, respectively), high field SC magnets (16T, 16T and 12 T, respectively), the total AC site power of 200 MW (HE-LHC) to $\sim$500 MW and cost 
7.2 BCHF (HE-LHC) to 17.1 BCHF (FCC-hh, assuming that 7 BCHF tunnel is available). In all these options, the detectors will need to operate at luminosities of $O(10^{35}$ cm$^{-2}$s$^{-1}$) and corresponding pileup will be $O$(500). 

Serious R\&D program that might take 12-18 years is needed to address most critical technical issues, such as development of accelerator quality 16 T dipole magnets based on Nb$_3$Sn (or 12 T iron-based HTS magnets for the SppC); effective intercept of the synchrotron radiation (5 MW in FCC-hh and 1 MW in SppC); beam halo collimation with circulating beam power 7 times that of the LHC; choice of optimal injector (eg., 1.3TeV scSPS, or 3.3 TeV ring either in the LHC tunnel or the FCC tunnel); overall machine design issues (IRs, pileup, vacuum, etc); power and cost reduction, etc. To be noted that such machines can also be used for ion-ion/ion-proton collisions; also high energy proton beams can be collided with high intensity $O$(60) GeV electrons out of an ERL (Input 159). 
\begin{figure}[h]
\centering
\includegraphics[width=80mm]{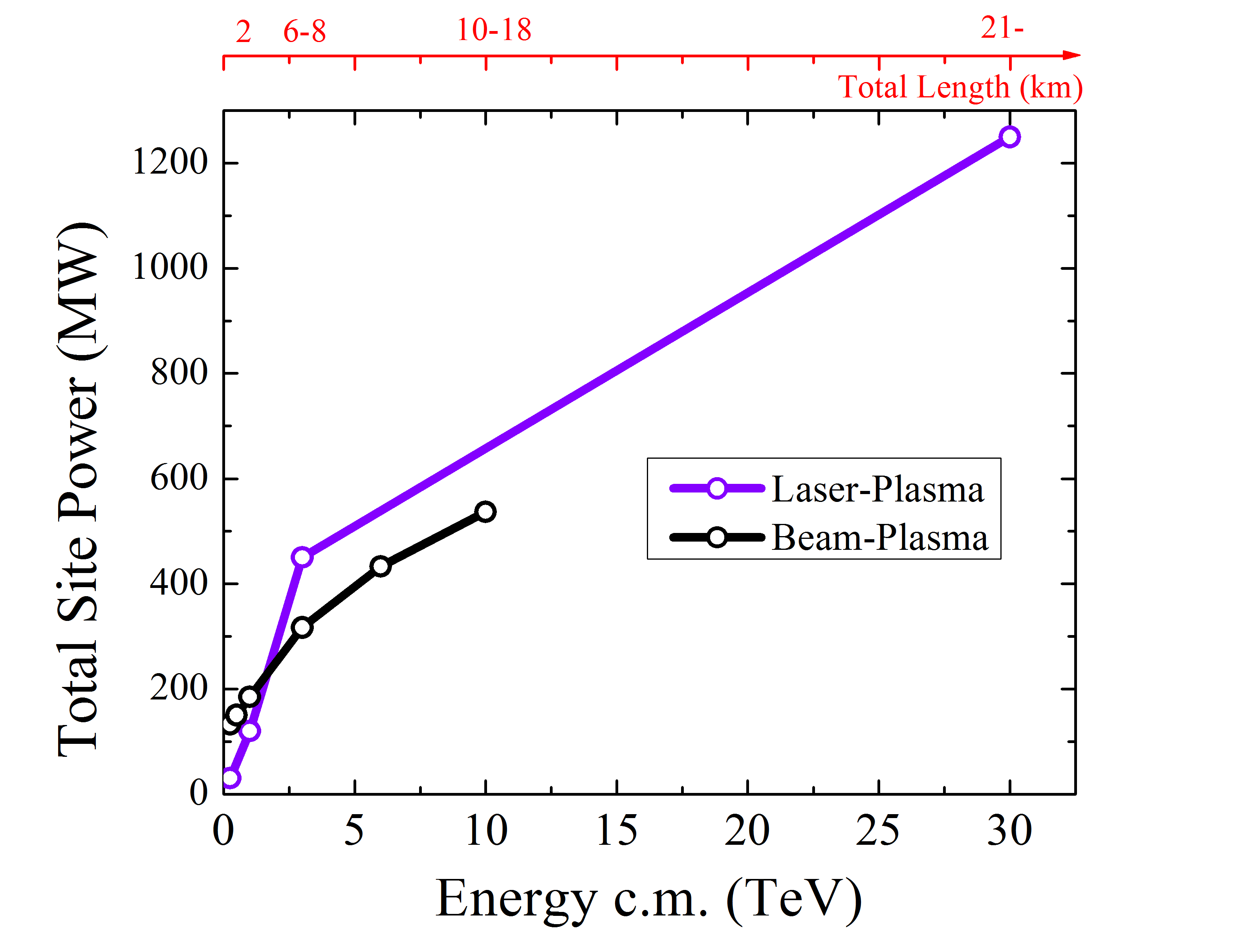}
\caption{Total length (upper scale) and total AC power requirements of multi-TeV $e^+e^-$ collider schemes based on plasma wakefield acceleration (data from the EPPSU Input 7). } 
\label{PWApower}
\end{figure}

Muon colliders offer in some aspects moderately conservative, in others - moderately innovative path to cost affordable energy frontier colliders. Again, their major advantages are a) muons do not radiate as readily as electrons and not affected by the beamstrahlung that allows efficient acceleration in rings at low cost and with great power efficiency; and b) the energy reach in muon collisions is about factor of 7 compared to the same energy $pp$ collisions - see Fig.\ref{MuonProton}. 

The muon collider R\&D had a number of remarkable advances in the past decade: the ionization cooling of muons was demonstrated in the MICE experiment at RAL (4D emittance reduction by $O$(10\%)); NC RF gradients 50 MV/m obtained in 3 T fields at Fermilab; also at Fermilab - rapid cycling HTS magnets achieved record ramping rate of 12 T/s; the first RF acceleration of muons demonstrated at JPARC MUSE RFQ (90 KeV); no cooling low emittance LEMMA concept is proposed (to use 45 GeV positrons to generate muon pairs at threshold). The {\it US MAP Collaboration} and its international partners have successfully carried out feasibility studies and showed that muon colliders can be built with the present day SC magnet and RF technologies; there is a well-defined path forward and initial designs exist for 1.5 TeV, 3 TeV, 6 TeV and 14 TeV colliders (the later, in the LHC tunnel and with the US MAP type proton driver\cite{Neuffer} is marked by asterisk in the Table \ref{future}). Still, there are many remaining issues to resolve which call for test facilities to demonstrate effective muon production and 6D cooling, to study the LEMMA scheme \cite{LEMMA} as well as to further develop concepts of fast acceleration, to deal with high detector background rates and the neutrino radiation issue. 

Last but not least, one has to mention impressive progress of new methods of
acceleration in plasmas (Inputs 7, 109, 58. 95). There are three ways to excite plasma wake-fields: by lasers (demonstrated electron energy gain of about 8 GeV over 20 cm of plasma with density 3$\cdot$10$^{17}$ cm$^{-3}$ at the BELLA facility in LBNL); by short electron bunches (9 GeV gain over 1.3m $\sim$10$^{17}$ cm$^{-3}$) plasma at FACET facility in SLAC) and by proton bunch (some 2 GeV gain over 10 m of 10$^{15}$ cm$^{-3}$ plasma at AWAKE experiment at CERN). In principle, the method of plasma wake field acceleration (PWFA) can make possible multi-TeV  $e^+e^-$ colliders. There is a number of critical issues to resolve along that paths: power efficiency of the laser/beam PWFA schemes; acceleration of positrons (which are defocused when accelerated in plasma); efficiency of staging (beam transfer and matching from one short plasma accelerator cell to another);  beam emittance control in scattering media; beamstrahlung (that leads to the rms energy spread at IP of about 30\% for 10 TeV machines and 80\% for 30 TeV collider - see Fig.\ref{PWApower}). The last four of these issues can possibly be addressed by accelerating muons (instead of electrons) in crystals or carbon nanotubes (density $\sim$10$^{22}$ cm$^{-3}$) –- the maximum theoretical accelerating gradient scales as square root of the plasma and can reach 1-10 TeV/m  allowing to envision a compact 1 PeV linear crystal muon collider \cite{VS2012}, \cite{2019WRKSH}. 

The PWFA technology is being actively explored becuase of  the opportunities it offers for a number of non-HEP applications. There are several proposals of plasma-based electron injectors for operational circular machines (a 100 MeV injector to the IOTA ring at  FNAL \cite{IOTA}; a 700 MeV injector to the PETRA-IV booster in DESY). Several collaborations are formed worldwide (EuPRAXIA, ALEGRO study, ATHENA) and facilities built or being built (in Europe - PWASC, ELBE/HZDR, AWAKE, CILEX, CLARA and SCAPA, EuPRAXIA at SPARCLAB at INFN-LNF, Lund, JuSPARC at FZJ and FLASHForward and SINBAD at DESY; also ImPACT in Japan , SECUF in China and FACET-II and BELLA in the US). Roadmap of the advanced acceleration concepts R\&D in the US aims at the PWFA collider CDR by 2035.

\section{Summary}
At present, aspirations of the high energy particle physics community include future Higgs factories and the Energy Frontier colliders. There are four feasible concepts: linear $e^+e^-$ colliders, circular $e^+e^-$ colliders, $pp/ep$ colliders and muon colliders. They all have limitations in energy, luminosity, efficiency and cost. The most critical (“Tier 0”) requirement for a Higgs factory is high luminosity and there are four proposals which generally satisfy it: ILC, CLIC-0.38, CEPC and FCC-ee. Te next level “Tier I” criteria include (in order): facility cost, the required AC power and readiness. The construction cost, if calibrated to performance (i.e., in GCHF/ab$^{-1}$) is the lowest for the FCC-ee, followed by CepC ($\times$4), then ILC (another $\times$10), then CLIC (another $\times$2) - see Table \ref{future}. The required AC site power consumption, if calibrated to performance (i.e., in TWh/ab$^{-1}$) is the lowest for the FCC-ee, followed by CepC ($\times$2), then ILC (another $\times$2), then CLIC (another $\times$2). As for the readiness, the ILC as a project is somewhat ahead of other proposals (it has TDR vs CDRs for CLIC, CepC, and FCC-ee) and its technical readiness is quite matured and includes industrial participation. The FCC-ee and CEPC proposals are based on the concepts and beam dynamics parameters that have already been proven at many past and presently operating circular colliders. 

The most critical “Tier 0” requirement for the energy frontier colliders is the center-of-mass  energy reach and there are four proposals which generally satisfy it (in order of the higher energy reach): CLIC-3 TeV, HE-LHC, 6/14 TeV Muon Collider and FCC-hh/SppC. The next level “Tier I” criteria for the EF machines are (in order); cost, AC power and R\&D effort. The construction cost is the lowest for the HE-LHC and the Muon Collider, followed by CLIC-3 ($\times$2) and  FCC-hh (another $\times$1.5). The required AC site power consumption, is the lowest for the HE-LHC and the Muon Collider, followed by CLIC ($\times$2), then by FCC-hh (another $\times$1.5). As for the required duration/scale of the R\&D effort to reach the TDR level of readiness, the CLIC-3 project is ahead of other proposals as it requires $\sim$10 years of R\&D vs about twice that for the HE-LHC and FCC-hh/SppC, and for the Muon Collider (the latter being at present the only concept without a comprehensive CDR). 

Arguably the hardest challenge for the EF hadron and muon colliders is development of representing magnets with maximum 16T field. There are fundamental challenges in getting the required current density in superconductors and in dealing with the ultimate magnetic pressures and mechanical stresses in the superconductor and associated components. Experts estimate that 20 to 30 years might be needed to innovate new approaches and technology to overcome the above-mentioned limits through continuous R\&D efforts. Lowering the maximum field requirement to 12-14 T or even to 6-9 T can greatly reduce the time needed for short-model R\&D, prototyping and pre-series work with industry. To realize even higher fields - beyond 16 T, if needed - High Temperature Superconductor (HTS) technology will be inevitably required. At present, the most critical constraint for the HTS is its much higher cost, even compared with the Nb$_3$Sn superconductor. 

Impressive advances of the exploratory PWFA R\&D over the past decade make it important to find out whether a feasible "far future" lepton collider option for particle physics can be based on that technology. One should note that PWFA has potential for non-HEP applications and has drawn significant interest and support from broader community, most notably, because of its possible use for X-ray production. Several research and test facilities are already built and operated, and many more are being planned. It will be important for the HEP accelerator designers to learn from the corresponding experience, understand applicability of the PWFA advances for particle colliders and encourage further technological development of the method.

\begin{acknowledgments}
This presentation with slight modifications was also given at the European Particle Physics Strategy Update Symposium (May 13-16, 2019, Granada, Spain). 
I would like to acknowledge input from and fruitful discussions on the subject of this presentation and thank M.Benedikt (CERN), P.Bhat (FNAL), M.Benedikt (CERN), C.Biscari (ALBA), A.Blondel (CERN), J.Brau (UO), O.Bruning(CERN), A.Canepa (FNAL), W.Chou (IHEP, China), M.Klein (CERN), J.P.Delahaye (CERN), D.Denisov (BNL), V.Dolgashev (SLAC), E.Gschwendtner (CERN), A.Grasselino (FNAL), W.Leemans
(DESY), E.Levichev (BINP), B.List (DESY), H.Montgomery (JLab), P.Muggli (MPG),
D.Neuffer (FNAL), H.Padamsee (Cornell), M.Palmer (BNL), N.Pastrone (INFN),
Q.Qin (IHEP), T.Raubenheimer (SLAC), L.Rivkin (EPFL/PSI), A.Romanenko
(FNAL), M.Ross (SLAC), D.Schulte (CERN), A.Seryi (Jlab), T.Sen (FNAL), F.Willeke
(BNL), V.Yakovlev, A.Yamomoto (KEK), F.Zimmermann (CERN), A.Zlobin (FNAL).   

Fermi National Accelerator Laboratory is operated by Fermi Research Alliance, LLC under Contract No. DE-AC02-07CH11359 with the
United States Department of Energy.

\end{acknowledgments}

\bigskip 

\end{document}